\documentclass[12pt]{article}
\usepackage{graphicx}
\usepackage{amssymb,amsmath,amsfonts,palatino,amsthm,epsf,epsfig}
\newcommand{\beq}{\begin{equation}}
\newcommand{\eeq}{\end{equation}}
\newcommand{\barr}{\begin{array}}
\newcommand{\earr}{\end{array}}

\def\one{{\mathchoice{\rm 1\mskip-4mu l}{\rm 1\mskip-4mu l}{\rm 1\mskip-4.5mu l}{\rm
1\mskip-5mu l}}}
\def\ket#1{| #1 \rangle}

\def\kb#1#2{|#1\rangle\!\langle #2 |}
\def\II{1\!\mathrm{l}}
\def\A{{\cal A}}
\def\B{{\cal B}}

\def\H{{\cal H}}
\def\K{{\cal K}}
\def\M{{\cal M}}
\def\P{{\mathcal P}}

\def\S{{\cal S}}
\def\U{{\cal U}}
\def\V{{\cal V}}

\def\Tr{{\mathrm{Tr}}}

\newcommand{\qforal}{\quad\text{for all}\quad}

\newtheorem{theorem}{Theorem}

\newtheorem{definition}{Definition}

\begin{document}
\title{Geometry from quantum particles}

\author{David~W.~Kribs\thanks{Email address: dkribs@uoguelph.ca}\\
Department of Mathematics and Statistics,\\
 University
of Guelph, Guelph, ON, Canada, N1G 2W1, and \\  Institute for
Quantum Computing,\\
 University of Waterloo, ON Canada, N2L 3G1\\
and\\
Fotini Markopoulou\thanks{Email address: fotini@perimeterinstitute.ca}\\
Perimeter Institute for Theoretical Physics,\\
35 King Street North, Waterloo, Ontario N2J 2W9, Canada, and \\
Department of Physics, University of Waterloo,\\
Waterloo, Ontario N2L 3G1, Canada}

\date{October 10, 2005}
\maketitle
\vfill
\begin{abstract}
We investigate the possibility that a background independent quantum theory of gravity is not  a theory of quantum geometry.  We provide a way for  global spacetime symmetries to emerge from a background independent theory without geometry.  In this, we use a quantum information theoretic formulation of quantum gravity and the method of noiseless subsystems in quantum error correction.  This is also a method that can extract particles from a quantum geometric theory such as a spin foam model.
\end{abstract}
\vfill
\newpage

\section{Introduction}

Is quantum gravity a theory of  quantum geometry, or is spacetime only a classical concept?

Most developed background-independent approaches to quantum gravity (i.e., theories whose basic quantities do not refer to a fixed spacetime geometry), such as loop quantum gravity \cite{LQG}, causal sets \cite{CauSet}, spin foams \cite{SF}, quantum Regge calculus \cite{QRC} or causal dynamical triangulations \cite{CDT} provide a candidate for a fundamental microscopic structure of spacetime that is some kind of quantum geometry: the kinematical state space of loop quantum gravity is a quantum superposition of spatial geometry states, causal sets is a path-integral of discrete causal orders, etc.   The goals of such theories are to:  i) be a well-defined microscopic theory of quantum geometry, ii) show that general relativity (and possibly also quantum field theory or matter couplings) emerge as the low-energy limit of the theory, and iii) make predictions on the kind and magnitude of departure from the classical theory.

These are very ambitious goals and each theory has had various levels of success so far.   Several of the difficulties these approaches face  (with the notable exception of causal dynamical triangulations) can be traced to the apparently desirable feature of background-independence and resulting complications, especially for the dynamical part of the theory.  At the same time, it has been suggested a number of times that the quantum theory of gravity may not be a quantization of general relativity (for example in \cite{Jac}).  While this suggestion has been supported by intriguing arguments, there are, so far, few suggestions on what the alternative may be.  Furthermore, if this is so, does one need a whole new type of background-independent quantum gravity formalism, or can at least some of the known approaches be modified to fit an expanded use (presumably revising the meaning of background-independence) in which quantum geometry can only be classical?

What could possibly be a way for spacetime to be a classical concept only, yet emergent from a background-independent quantum theory?  A possible answer is to start from a quantum, suitably background-independent theory and first look for long-range coherent degrees of freedom.  These will characterize the low-energy limit.  They can be thought of as particles even though, at this level, there is no spacetime and thus the usual notion of particles (as in Wigner) does not apply.  Only then, if these behave as if they are in a spacetime, do we have a spacetime.

How can one make sense of this scenario?  On the face of it, there is a problem with every single step in the above.  First, what is a {\em long-range} propagating degree of freedom if there is no spacetime and thus no obvious notion of ``long-range''?  What can we mean by ``low-energy limit'', when energy needs time-translation invariance?  What can a particle be in this setup?  And, if one can define it, what does it mean that ``it behaves as if in a spacetime''?

The aim of this paper is to investigate how far one can go in this direction by following \cite{M00B} that suggested that unnecessary references to spacetime can be eliminated by using the language of quantum information processing.  Indeed,  an object such as a spin foam can be formulated as a quantum superposition of quantum information flows which then may be restricted to be geometric and thus reduced to the usual quantum geometry path integrals, or not, the case of interest for us.

In this setup, as suggested by \cite{MarPou}, a suitable notion of
a particle from quantum information processing can be used.  This
is the notion of a noiseless subsystem in quantum error
correction, a subsystem protected from the noise, usually thanks
to symmetries of the noise.  In a quantum gravity context, this
means a subsystem emergent (protected) from the microscopic
Planckian  evolution, and thus relevant for the effective theory.
The extent to which this is background-independent will be
discussed in detail in the next section and the concluding one. We
will then suggest that these ``behave as if they are in a
spacetime'' if they are invariant under Poincar\'{e}
transformations, or deSitter in the case of a positive
cosmological constant, and so do their interactions.  That is, we
turn around the usual order:  a particle is not Poincar\'{e}
invariant because it is in a Minkowski spacetime, rather, all we
can mean by a Minkowski spacetime is that all coherent degrees of
freedom and their interactions  are Poincar\'{e} invariant at the
relevant scale.

In a simplified setup (where all coherent degrees of freedom are free and the relevant transformations are not emergent, as one expects in the full theory, but already present in the microscopic dynamics) we provide the required conditions on the fundamental dynamics that realize this scheme.

A secondary goal of this paper is to address the low energy problem in background independent approaches to quantum gravity, namely the problem of extracting a semiclassical low energy geometry from a dynamical microscopic quantum geometry.   That is, our results may also be useful to quantum theories of gravity {\em with} microscopic quantum geometry:  the definition of a coherent degree of freedom we use can be applied, for example, to spin foams with a boundary to extract its effective particles (and, in fact, that is why it was originally considered in \cite{MarPou}).  Our setup thus provides a new way to get to the much sought-after semiclassical limit.    In future work, we hope to give an
algorithmic construction of the class of microscopic dynamics that contains Poincar\'{e}-invariant particles.

The outline of this paper is as follows.  In the next section we
clarify what we mean by background independence. In section 3, we
review how a quantum gravity model can be described as a quantum
information processing system.  In section 4,   we expand on the
suggestion of \cite{MarPou} to use quantum error correction to
identify long-range propagating degrees of freedom.  We review the
particular kind of quantum error correction that is relevant here:
decoherence-free subspaces and noiseless subsystems. These
coherent subsystems are required to be invariant under certain
unitary transformations (Poincar\'{e}, Lorentz or deSitter would
be an example) which implies further specific conditions on the
fundamental dynamics.   We discuss the strengths, weaknesses and
implications of this setup in the Conclusions.

\section{Background-independent quantum gravity}

General relativity tells us that only events and their relations
are physical.  Any coordinates we may use to describe them have no
physical meaning and coordinate distances are not physical
quantities. In general relativity the metric is dynamical: there
is no background space and time. Possibly the most concise
statement of background-independence is the one given by Stachel:
``There is no kinematics independent of dynamics'' \cite{Sta}.
Since the quantum theory of gravity is expected to contain general
relativity, it is reasonable to ask that it maintains this
important principle and is also background independent.  Of
course, we do not know what form background-independence will
ultimately take, if it applies at all.  There is a substantial
literature discussing different possible forms of background
independence (see \cite{Sta, ButIsh, Smo} and references therein).

We are interested in a  background independent quantum theory of gravity which is not based on quantum geometry.  In that context, we note that in the literature and in quantum gravity research one finds two statements of background independence which lead to very different conclusions relevant for us.  They can be summarized as follows \cite{ButIsh}:

{\begin{description}
\item[Background independence 1.]
A quantum  theory of gravity is background independent if its basic quantities and concepts do not presuppose the existence of a given background metric.

\item[Background independence 2.]
A  quantum theory of gravity is background independent if there is no fixed theoretical structure.  Any fixed structure will be regarded as a background.
\end{description}

The main background independent approaches to quantum gravity implement both 1 and 2 since they are given as theories of dynamical quantum geometry.  That is, the two statements cannot be distinguished.
Although there is no implementation of 2 outside the quantum geometry context, we feel it is important that it should be distinguished from 1.

It is clear that a quantum theory of gravity in which the basic quantities are not geometric in any sense ought to be automatically background-independent in the sense of 1, whether or not there is some fixed structure in it.  This will be our position for the purposes of this article.

In a little more detail, since one of the main background independent quantum gravity candidates, loop quantum gravity, is a canonical theory, we would like to clarify the following.  Often in the literature, a merging of 1 and 2 above, appears to imply that  the fundamental quantum theory of gravity must have a quantum Hamiltonian constraint and that a true Hamiltonian evolution in the microscopic theory is excluded.    Our viewpoint is that this applies to (globally hyperbolic) quantum geometry and needs not be imposed on a candidate quantum theory of gravity in which geometry is only classical (in this we follow \cite{Llo} and \cite{Dre}).  Such a theory, instead, only needs the classical Hamiltonian constraint to be present at the regime at which classical geometry arises\footnote{A common objection to a microscopic Hamiltonian that surely a true Hamiltonian must imply a preferred time, which is then ruled out by observations.  This is not necessarily the case.  First, there is no reason that the evolution of the fundamental quantum degrees of freedom has a direct correspondence to the geometric spacetime description. Second, even classically, one can have, for example,  multifingered evolution with a fixed average lapse.\\
Also note that one can be perfectly relational without resorting to the extreme of the second form of background independence.  All that is required is that any physically relevant observable refers to observers inside the system and their relations.}.

In short, our position in this paper is that to impose the second form of background independence, whether or not the relevant quantum quantities are geometric,  is to make an arbitrary extrapolation from the background-independence of general relativity.   As for the first form, since its implementation in terms of quantum geometry
 creates a lot of problems, especially in the key issue of the low-energy limit of a microscopic quantum theory of gravity, we will not adopt it.  Instead, we investigate the possible role of classical geometry in a quantum theory of gravity whose microscopic degrees of freedom are not geometric.  This surely is background independent in the sense of 1.

\section{Background-independent quantum gravity as a \\
quantum information processing system}

The type of quantum theory of gravity we will consider for the
purposes of our work is a {\em quantum causal history}
\cite{M00B,HMS03,M02,M00C,M00A,G00,MS98,MS97}. Mathematically, a
quantum causal history is a directed graph with a
finite-dimensional Hilbert space associated with each vertex and a
quantum channel associated with each edge (details follow). This
formalism is of interest because it can be adapted to fit two
different roles:

1.  At the mathematical level, with no reference to any geometric properties of the graph and the edges, this is simply a quantum information processing system. This is (trivially) background-independent in the sense discussed in the previous section.  In this form, it is suitable for our intended application of a quantum information theoretic method that introduces geometrical properties at the level of effective coherent degrees of freedom encoded in the system.  The aim will be to find global symmetries of a classical geometry at the level of particles, without starting with a quantum geometry.

2.   It is also possible to link this structure to quantum
geometry via the interpretation of the directed graph as a causal
set and the vertices as the events.  Dynamics is introduced via a
path-integral sum over all causal sets interpolating between given
in and out states. This gives a spin foam model and thus a quantum
geometry \cite{M00B}.  In fact, quantum causal histories were
introduced in \cite{M00B} and further developed in \cite{HMS03} as a
formalism for a quantum cosmology which is locally finite, causal
and background-independent.  In this case, the same quantum
information theoretic method can identify effective Poincar\'{e}
particles in the spin foam, thus providing a new way to
investigate the semiclassical limit of such models.

In the remainder of this section we review the basics of quantum
causal histories, both in its ``bare bones'' version and the
quantum geometric one.


\subsection{Quantum causal histories on directed graphs}

Let $\Gamma$ be a directed graph with vertices $x\in V(\Gamma)$
and directed edges $e\in E(\Gamma)$. The {\it source} $s(e)$  and
{\it range} $r(e)$ of an edge $e$ are, respectively, the initial
and final vertices of $e$. A (finite) path $w=e_k \cdots e_1$ in
$\Gamma$ is a word in the edges of $\Gamma$ such that $r(e_i) =
s(e_{i+1})$ for $1\leq i < k$. If $s(w) = r(w)$ then we say $w$ is
a {\it cycle}. We require that $\Gamma$ has no
cycles\footnote{Note that, while this condition was initially
motivated by $\Gamma$ being a causal set (see \ref{qchC}), the
same condition is also natural if the quantum causal history is a
quantum computer with $\Gamma$ the circuit.}. If there exists a
path $w$ such that   $s(w)=x$ and $r(w)=y$
 let us write $x\leq y$ for
the associated partial ordering.  We call such vertices {\em related}.  Otherwise, they are {\em unrelated} and we use $x\sim y$ to denote this.  Given any $x,y\in V(\Gamma)$ there are finitely many $z\in V(\Gamma)$
such that $x\leq z \leq y$.

A {\it quantum causal history}  is a directed graph $\Gamma$
endowed with the following structure.
For every
vertex $x\in V(\Gamma)$ there is a (finite-dimensional) Hilbert space
$\H(x)$.  If $x$ and $y$ are unrelated,   the joint Hilbert space is $\H(x)\otimes H(y)$.

For every edge $e\in E(\Gamma)$ there is a quantum channel
\beq
\Phi_e: {\cal A}(s(e))\rightarrow{\cal A}(r(e)),\label{eq:qchphi}
\eeq
where ${\cal A}(x)$ is the full matrix algebra on $\H(x)$. (Basic
properties of quantum channels are discussed below.)
Given $x\neq y$, without loss of
generality we can assume there is at most one directed edge from
$x$ to $y$. If there are multiple edges then the corresponding
channels can be combined into one channel which contains the
pertinent information in the quantum causal history.

A {\em parallel set} $\xi\subseteq E(\Gamma)$ is defined by the
property that $x\sim y$ whenever $x,y\in\xi$. The algebra ${\cal
A}(\xi) = \otimes_{x\in\xi} {\cal A}(x)$ acts on the composite
system Hilbert space ${\cal H}(\xi) = \otimes_{x\in\xi} {\cal
H}(x)$.  If $\xi$ and $\zeta$ are two parallel sets such that all
forward directed paths from $\xi$ intersect $\zeta$ and all paths
that arrive at $\zeta$ pass through $\xi$, then we have an
evolution of a closed quantum system and a unitary operator \beq
U\left(\xi,\zeta\right):{\cal H}(\xi)\rightarrow {\cal H}(\zeta).
\eeq This determines an isomorphism $\Phi(\xi, \zeta):{\cal
A}(\xi)\rightarrow{\cal A}(\zeta)$ via \beq
\Phi\left(\xi,\zeta\right)\left(\rho\right)=U\left(\xi,\zeta\right)\rho
\,
    U\left(\xi,\zeta\right)^\dagger\label{eq:qchU}\qforal\rho\in{\cal A}(\xi).
\eeq Eq.\ (\ref{eq:qchphi}) is the restriction of (\ref{eq:qchU})
to ${\cal A}(x)\subseteq{\cal A}(\xi)$ for $x\in\xi$.  In turn,
(\ref{eq:qchU}) can be reconstructed from the local maps
(\ref{eq:qchphi}) using the appropriate precise mathematical
definition of a quantum causal history (see \cite{HMS03} for more
details).

At this level, the quantum causal history is simply a quantum
information processing structure. Note that it is possible to
mirror the original quantum geometric construction by introducing
a ``path-integral'' superposition of all possible graphs
interpolating between two given sets of commuting algebras.
However, there is no motivation for doing so if $\Gamma$ is an
information flow circuit and not a discrete quantum geometry.  In
any case, our results apply to both cases.

\subsection{Quantum Channels}\label{S:channels}

Quantum channels are central both in quantum causal histories and in the method we will use to extract particles from them.  Here we give a brief account of their basic features.

Let $\H_S$ be the state space of a quantum system in contact with
an environment $\H_E$. The standard characterization of evolution
in open quantum systems starts with an initial state in the system
space that, together with the state of the environment, undergoes
a unitary evolution determined by a Hamiltonian on the composite
Hilbert space $\H = \H_S \otimes \H_E$, and this is followed by
tracing out the environment to obtain the final state of the
system. The associated evolution map, or ``superoperator'',
$\Phi:\B(\H_S)\rightarrow\B(\H_S)$ is necessarily completely
positive (see below) and trace preserving. More generally, the map
could have different domain and range Hilbert spaces. Hence the
operational definition of a {\it quantum channel} (or quantum
evolution, or quantum operation) from a Hilbert space $\H_1$ to
$\H_2$, is a completely positive, trace preserving map $\Phi:
\B(\H_1) \rightarrow \B(\H_2)$.

A {\it completely positive} (CP) map  $\Phi$ is a linear map $\Phi
: \B(\H_1)\rightarrow\B(\H_2)$ such that the maps
\[
 id_k \otimes \Phi :  \M_k \otimes \B(\H_1) \rightarrow
 \M_k \otimes \B(\H_2)
\]
are positive for all $k\geq 1$. Here we have written $\M_k$ for
the algebra $\B(\mathbb{C}^k)$ represented as the $k\times k$
matrices with respect to a given orthonormal basis. (The CP
condition is independent of the basis that is used.)

A fundamental technical device in the study of CP maps is the {\it
operator-sum representation} given by the theorem of Choi
\cite{Cho75} and Kraus \cite{Kra71}. For every CP map $\Phi$ there
is a set of operators $\{E_a\}\subseteq \B(\H_1,\H_2)$ such that
\begin{eqnarray}\label{opsum}
\Phi(\rho) &=& \sum_a E_a \rho E_a^\dagger \qforal
\rho\in\B(\H_1).
\end{eqnarray}
We shall write $\Phi = \{E_a\}$ when the $E_a$ satisfy
Eq.~(\ref{opsum}) for $\Phi$. The family $\{E_a\}$ may be chosen
with cardinality $|\{E_a\}|\leq \dim(\H_1)\dim(\H_2)$, and is
easily seen to be non-unique\footnote{However, if $\{E_a\}$ and
$\{F_b\}$ are two families of operators that implement the same
channel $\Phi$, then there is a scalar matrix $U=(u_{ab})$ such
that $E_a = \sum_b u_{ab} F_b$ for all $a$.}.

The class of CP maps that are quantum channels satisfy an extra
constraint. Specifically, note that when $\Phi$ is represented as
in (\ref{opsum}), trace preservation is equivalent to the identity
\begin{eqnarray}\label{tracepreserve}
 \sum_a E_a^\dagger  E_a = \one_{\H_1}.
\end{eqnarray}
Thus, a quantum channel $\Phi$ is a map which satisfies
(\ref{opsum}) and (\ref{tracepreserve}) for some set of operators
$\{E_a\}$.

\subsection{Quantum causal histories with geometrical information}\label{qchC}

The quantum causal history can also double as a formalism of a
microscopic quantum geometry theory. We start by interpreting the
directed graph $\Gamma$ as a causal set (a discrete, locally
finite analogue of the set of events of a Minkowski spacetime) and
the vertices as events. These are the smallest Planck scale
systems in a quantum spacetime. In a locally finite theory (i.e.,
with a finite number of relevant degrees of freedom in a finite
volume) these quantum systems are assigned a simple matrix algebra
$\A(x)$ for each event $x$. Two unrelated events are acausal, thus
the operators on the corresponding algebras commute. Every causal
relation $x\leq y$ and the corresponding edge $e$ in the causal
set is the evolution of an open quantum system and hence a quantum
channel $\phi_e:\A(x)\rightarrow \A(y)$.

A path integral model of a quantum spacetime is then given as a quantum sum over all possible causal sets that interpolate between two given ``parallel'' sets of events $S_i$ and $S_f$ (corresponding to the universe at a given initial and final times), formally:
\beq
A_{S_i\rightarrow S_f}=\sum_{\Gamma:S_i\rightarrow S_f}\prod_{e\in \Gamma}\phi_e.\label{eq:PI}
\eeq

One can import further geometric information, for example by
requiring that the local state spaces are the spin network
intertwiner spaces of loop quantum gravity, i.e., $\H(x)$ is the
vector space of so-called intertwiners.  These are maps from the
tensor product of representations of SU(2) to the identity
representation.  This, and other assignments of different groups
and intertwiners, are examples of {\em spin foam models} \cite{SF}.

A microscopic model of spacetime is successful if it has a good low-energy limit in which it reproduces the known theories, namely general relativity with quantum matter coupled to it.  In the case of causal dynamical triangulations, impressive results show strong indications that this model has the desired features \cite{CDT}.  This hinges on specific features of the model that allow a Wick rotation to a statistical sum and thus technical control of the sum via both analytic and numerical methods.  In the general spin foam case, this is a formidable problem, involving a quantum sum over group representations.

Our proposal regarding this problem is that, instead of considering the sum (\ref{eq:PI}) directly, one can first look for long-range propagating degrees of freedom (particles) and reconstruct the geometry from these (if they exist).  The specific method we adopt is promising because it deals directly with quantum systems and coarse-grains a quantum system to its effective particles.  Our discussion applies to such models with a boundary.

\section{Particle: Group-invariant noiseless subsystems}

We now suggest that a suitable definition of a coherent degree of
freedom in a quantum causal history is a {\em noiseless
subsystem}.

Quantum channels depict the most general form of evolution in open
quantum systems, and hence they play a central role in quantum
computing. In this setting, the operators $\Phi = \{E_a\}$ in the
operator-sum representation Eq.~(\ref{opsum}) for a channel are
called the {\em error} or {\em noise} operators associated with
$\Phi$. It is precisely the effects of such operators that must be
mitigated for in the context of quantum error correction.

The noiseless subsystem method (also called decoherence-free
subspaces and subsystems) is the fundamental passive technique for
error correction in quantum computing. Recently a framework for
studying noiseless subsystems that applies to arbitrary quantum
channels was presented \cite{KLP05,KLPL05,CK05}. This framework is
built upon earlier work in passive quantum error correction
\cite{PSE96,DG97c,ZR97c,LCW98a,KLV00a,Zan01b,KBLW01a}, and is a
centrepiece of the unified approach to quantum error correction,
called ``operator quantum error correction'', introduced in
\cite{KLP05,KLPL05}. The basic idea in this setting is to (when
possible) encode initial states in sectors that will remain immune
to the degrading effects of errors $\Phi = \{E_a\}$ associated
with a channel.

The mathematical description is given as follows. Let $\Phi$ be a
channel on $\H$ and suppose that $\H$ decomposes as $\H =
(\H^A\otimes\H^B)\oplus\K$, where $A$ and $B$ are subsystems and
$\K =(\H^A\otimes\H^B)^\perp$. We say that $B$ is {\em noiseless}
for $\Phi$ if
\begin{eqnarray}\label{ns}
\forall\sigma^A\ \forall\sigma^B,\ \exists \tau^A\ :\
\Phi(\sigma^A\otimes\sigma^B) = \tau^A\otimes \sigma^B.
\end{eqnarray}
Here we have written $\sigma^A$ (resp. $\sigma^B$) for operators
on $\H^A$ (resp. $\H^B$), and we regard $\sigma = \sigma^A\otimes
\sigma^B$ as an operator that acts on $\H$ by defining it to be
zero on $\K$. Let $P^{AB}$ be the projection of $\H$ onto
$\H^A\otimes\H^B$ and define a ``compression superoperator''
$\P^{AB}(\cdot) = P^{AB}(\cdot)P^{AB}$ on $\H$. That is, $\P^{AB}$
is the map on $\B(\H)$ defined by $\P^{AB}(\sigma) = P^{AB}\sigma
P^{AB}$, $\forall\sigma\in\B(\H)$. Then in terms of the partial
trace operation on $A$, Eq.~(\ref{ns}) is equivalent to the
statement
\begin{eqnarray}\label{ns1}
\Tr_A \circ \P^{AB} \circ \Phi \circ \P^{AB} = \Tr_A \circ
\P^{AB}.
\end{eqnarray}
See the Appendix for an example of a noiseless subsystem.  One
would like to eventually generalize this method to a suitable
notion of approximate noiseless subsystems.




We now attempt to identify what it should mean for a quantum
causal history to (i) contain a subsystem that evolves in a well
defined unitary manner and (ii) for this notion to be invariant in
a group-theoretic sense.

Let $G$ be a group and suppose that $\pi:G\rightarrow\B(\H^{\rm
rep})$ is a (unitary) representation of $G$ on a Hilbert space
$\H^{\rm rep}$. We identify $G$ with the unitary group $\pi(G)$.
For each $U$ in $G$ denote the corresponding superoperator on
$\B(\H)$ by $\U(\cdot) = U(\cdot)U^\dagger$. To further simplify
the notation below, we shall denote the representation Hilbert
space as $\H^B\equiv\H^{\rm rep}$. We are interested in scenarios
for which $\H^B$ is a subsystem of a larger Hilbert space $\H$
within a causal history. Specifically, $\H$ decomposes as $\H =
(\H^A\otimes\H^B)\oplus\K$, where $\K = (\H^A\otimes\H^B)^\perp$.
Suppose now that we have a quantum channel $\Phi$ on $\H$.



\begin{definition}
{\rm We say that $B$ is {\it $G$-invariant under $\Phi$} if there
is a $U$ in $G$ such that
\begin{eqnarray}\label{maineqn}
\forall\sigma^A\ \forall\sigma^B,\ \exists \tau^A\ :\
\Phi(\sigma^A\otimes\sigma^B) = \tau^A\otimes \U(\sigma^B).
\end{eqnarray}
}
\end{definition}

This terminology is justified in the following sense. As a
consequence of the theorem below, observe that Eq.~(\ref{maineqn})
may be restated as
\begin{eqnarray}\label{secondeqn}
\big( \Tr_A \circ \Phi \circ \iota_B \big) (\sigma^B) =
\U(\sigma^B)
 \quad \forall \sigma^B,
\end{eqnarray}
where $\iota_B:\B(\H^B)\rightarrow\B(\H)$ is the map given by
$\iota_B(\sigma^B) = \alpha(\one^A \otimes \sigma^B)$ with $\alpha
= ({\rm dim}\, \H^A)^{-1}$. Notice in particular that this implies
\begin{eqnarray}\label{inveqn}
\V\circ \Tr_A \circ \Phi \circ \iota_B = \V\circ \U  \quad \forall
\,\, V\in G.
\end{eqnarray}
The map $\V\circ\U$ is implemented by the unitary $VU\in G$. It
follows that evolution of the $B$ subsystem under $\Phi$ is
invariant for the natural group action of $G$.

In the context of a quantum causal history, this means that the
microscopic evolution in the history: i) leaves $\H^B$ to evolve
unitarily, i.e., it is an effective coherent degree of freedom and
ii) $\H^B$ is invariant under $G$-transformations, where the
interesting implementations are when $G$ is the Poincar\'{e},
Lorentz or deSitter group. Also note that the formulation of
Eq.~(\ref{maineqn}) can be widely applied within the causal
history framework. In particular, it could be applied to a single
edge map $\phi_e$, to a map $\phi_w =
\phi_{e_k}\circ\ldots\circ\phi_{e_1}$ associated with a path
$w=e_k\cdots e_1$, or, indeed, to any evolution map associated
with the structure of the history, such as partial traces over any
given subsystem, etc.

There is some notational clarification required for
Eq.~(\ref{maineqn}). The channels $\phi_e:
\B(\H(s(e)))\rightarrow\B(\H(r(e)))$ associated with a quantum
causal history map between different Hilbert spaces. This would
appear to be problematic in connection with Eq.~(\ref{maineqn}),
as the operator $U\sigma^B U^\dagger$ acts on the subsystem $\H^B$
of, in this case, $\H(s(e))$, and not of $\H(r(e))$. However, our
formulation of Eq.~(\ref{maineqn}) is simply a notational
convenience. We could identify $\H^B$ with a subsystem $\H^D$ of
$\H(r(e))$ via a unitary $V^{BD}:\H^B\rightarrow\H^D$ (or more
generally an isometry), and under this map $\sigma^B$ is
identified with $\sigma^D = V^{BD} \sigma^B (V^{BD})^\dagger$ and
the group element $U$ is identified with $U' = V^{BD} U
(V^{BD})^\dagger$. In the case of a quantum computer, for
instance, this could simply be the identification of the standard
basis for $n$-qubit space with itself, after a certain time step.
Thus, for brevity, we have suppressed this notational issue,
effectively assuming the map $V^{BD}$ is the identity map. With
this identification in mind, we can write Eq.~(\ref{maineqn})
unambiguously.

We note that Eq.~(\ref{maineqn}) first appeared in \cite{KLPL05}
in the context of operator quantum error correction as a natural
generalization of the notion of noiseless subsystems for quantum
operations introduced in \cite{KLP05,KLPL05}.

The following theorem gives a number of testable conditions that
are equivalent to $G$-invariance. Namely, the result shows how
this notion may be  phrased in terms of the partial trace
operation on $A$; that it is enough to satisfy this equation for
the maximally mixed state on $A$; how to test if a given subsystem
satisfies this equation if a choice of operator elements for the
evolution map is known; and the corresponding statement in terms
of operator algebras.

\begin{theorem}\label{thm:NS}
Let $G$ be a group represented on a Hilbert space $\H^B$. Suppose
that $\H$ is a Hilbert space that decomposes as $\H =
(\H^A\otimes\H^B) \oplus\K$, and that
$\Phi:\B(\H)\rightarrow\B(\H)$ is a quantum channel. Then the
following five conditions are equivalent:
\begin{itemize}
\item[1.] $B$ is $G$-invariant under $\Phi$. \item[2.] $\exists\,
U\in G : \forall\sigma^B,\ \exists \tau^A\ :\
\Phi(\one^A\otimes\sigma^B) = \tau^A \otimes \U( \sigma^B)$
\item[3.] $\exists\, U\in G :\ \Tr_A\circ \P^{AB}\circ \Phi\circ
\P^{AB} =\U\circ\Tr_A\circ\P^{AB}$. \item[4.] Let
$\{\ket{\alpha_k}\}$ be an orthonormal basis for $\H^A$ and let
$\{P_{kl} = \kb{\alpha_k}{\alpha_l}\otimes\one^B \}$ be the
corresponding family of matrix units in $\B(\H^A)\otimes\one^B$.
Let $\Phi=\{E_a\}$ be a choice of operator elements for $\Phi$.
Then there is a $U\in G$ such that
\begin{equation}
P_{kk} (\one^A\otimes U^\dagger) E_a P_{ll} = \lambda_{akl} P_{kl}
\quad\forall\, a,k,l \label{eq:cond1}
\end{equation}
for some set of scalars $\{\lambda_{akl}\}$ and
\begin{equation}
 (\one^A \otimes U^\dagger) E_a P^{AB} = P^{AB}  (\one^A \otimes U^\dagger)
 E_a P^{AB} \quad\forall\, a. \label{eq:cond2}
\end{equation}
\item[5.] There is a $U\in G$ such that the subspace
$\H^A\otimes\H^B$ is invariant for the operators $\{ (\one^A
\otimes U^\dagger) E_a \}$, and the restricted operators $\{
(\one^A \otimes U^\dagger) E_a P^{AB}\}$ belong to the operator
algebra $\B(\H^A)\otimes\one^B$.
\end{itemize}
\end{theorem}

{\noindent}{\it Sketch of proof.} Suppose that $B$ is
$G$-invariant under $\Phi$, and so Eq.~(\ref{maineqn}) is
satisfied for some $U\in G$. This is equivalent to the statement
that $\forall\sigma^A\ \forall\sigma^B,\ \exists \tau^A$ such that
\begin{eqnarray}\label{nseqn}
\ \big( (id_A\otimes\U^\dagger)
\circ\Phi\big)(\sigma^A\otimes\sigma^B) = \tau^A\otimes \sigma^B,
\end{eqnarray}
where $\U^\dagger(\cdot) = U^\dagger (\cdot) U$. Thus, in the
terminology of \cite{KLP05,KLPL05}, this shows that $\H^B$ is a
noiseless subsystem for the map $(id_A\otimes\U^\dagger )\circ
\Phi$. The result now follows from the characterization of
noiseless subsystems from \cite{KLP05,KLPL05}.  \hfill$\square$

Let us discuss a mathematical problem motivated by this
discussion. In the quantum causal history setting, we wish to
regard the group $G$ and a particular representation of $G$ as
given (namely, we know what the classical spacetime is and that
particles are representations of the Poincar\'{e} group). Thus, it
is of interest to find the evolution maps $\Phi$ such that
Eq.~(\ref{maineqn}) holds for some element $U$ of $G$. To be
precise, given $G$ and a representation of $G$ on a Hilbert space
$\H^{\rm rep}=\H^B$, find the set of all $\Phi$ such that $B$ is
$G$-invariant under $\Phi$. This is of interest as it would give a
class of microscopic quantum evolutions that contain particles
with the desired classical geometric properties.

This problem is also of interest in the context of error
correction in quantum computing. Indeed, in the case that $G$ is
the trivial group, this problem may be interpreted as follows:
Given a subsystem $\H^B$ of a system $\H$, find the quantum
operations $\Phi$ on $\H$ for which $\H^B$ is a noiseless
subsystem. The recent work \cite{CK05} solves a different, but
related problem. Specifically, an explicit method is presented to
compute all noiseless subsystems when $\Phi$ is given. We expect
that the techniques used in that work could lead to progress on
the problem described here.

\section{Conclusions: Are the Poincar\'{e} transformations the chicken or the egg?}

Our task in this paper was to investigate how a classical spacetime may be emergent from a background independent quantum theory of gravity whose basic quantities are not quantum geometric.  We proposed that the classical geometry can be present as symmetries of the emergent coherent degrees of freedom.  That is, we wish to understand how global spacetime symmetries can emerge in a background independent theory with no spacetime.   In parallel, we were interested in addressing the low energy issue of spin foam-like models via the use of emergent particles in the model.

We used a quantum information theoretic formulation and the specific method of noiseless subsystems used in the quantum error correction literature to characterize these coherent degrees of freedom, encoded in the microscopic dynamics.  We generalized this to a suitable notion of group-invariant noiseless subsystems, thus giving a condition for the theory to have the required global symmetries.  This opens up the exciting possibility of having an algorithmic construction of the class of microscopic dynamics that contains the desired encoded particles.

Let us note some of the most interesting features of noiseless
subsystems in a quantum gravity context.  First, they are not
localized, thus their symmetry is global  (see the example in the
Appendix).  This is also relevant to the discussion of microscopic
versus emergent locality in quantum gravity.  They illustrate the
fact that the emergent degrees of freedom can bear little relation
in their interactions to the underlying microscopic theory, known
of course from condensed matter physics, but now in a manifestly
background independent form. Second, the construction employs
quantum channels, rather than a partition function of the usual
spin foam type, which applies both to a single underlying circuit
(or history) or to a path integral sum.  Finally, it is very
important that the existence and properties of the noiseless
subsystems depends entirely on the properties of the dynamics.  As
can be seen in the quantum information literature
\cite{PSE96,DG97c,ZR97c,LCW98a,KLV00a,Zan01b,KBLW01a,KLP05,KLPL05,CK05}
and in the application of this method to quantum black holes
\cite{DMS04}, as well as the example in the Appendix,  in concrete
examples of noiseless subsystems their existence depends on having
symmetries in the dynamics.

It is also of interest that our results can be applied to spin
foams with a boundary to extract the particles they contain and
thus address the outstanding low energy issue of these models.
(The importance of the boundary is also emphasized in
\cite{Carlo}).

The following are shortcomings in the current application  of noiseless subsystems to quantum gravity and will need to be addressed in future work:   The $G$-invariant noiseless subsystems are not truly emergent but encoded in the microscopic dynamics, in the sense that both the symmetries of the dynamics that guarantee the existence of the noiseless subsystems and the group $G$ are present in the microscopic dynamics.  One would like to extend the relevant notions to an appropriate definition of an approximate $G$-invariant noiseless subsystem.  Also, in order to claim that there is a flat spacetime in the microscopic theory, we need to have G-invariant interactions between the noiseless subsystems.

\section*{Acknowledgments}

FM is grateful to several colleagues for extensive discussions on
this subject over the past two years, especially Olaf Dreyer, Eli
Hawkins, Tomasz Konopka, Seth Lloyd, David Poulin, Lee Smolin and
Paolo Zanardi. DWK would like to thank Raymond Laflamme, David
Poulin and Rob Spekkens for interesting discussions on related
topics.

\section*{Appendix:  An example of a noiseless subsystem in quantum computing}

We will now present a widely used example of a noiseless subsystem
in quantum computing. See \cite{HKL04,HKLP05,JKK05} for a detailed
analysis of related and more general noiseless subsystems. The
system $\S$ is composed of three spin-$\frac 12$ particles
(labeled $A$, $B$, and $C$), so $\H_\S = (\mathbb{C}^2)^{\otimes
3}$. The CP map is a collective rotation: all three spins get
rotated around a common axis and by a common angle, but these
specifications of the rotation are chosen at random and are
unknown. The rotation operator of a single spin-$\frac 12$
particle (an element of the Lie group $SU(2)$) can be written in
terms of the Pauli operators (the generators of the Lie algebra
$su(2)$): $\exp\{-i\theta\vec n\cdot\vec\sigma\}$ where $\vec n$
is a real three dimensional unit norm  vector defining the axis of
rotation, $\theta \in [0,2\pi]$ is the rotation angle, and $\vec
\sigma = (\sigma_x,\sigma_y,\sigma_z)$. Defining $\vec J =
\vec\sigma_A\otimes\II_B\otimes\II_C +
\II_A\otimes\vec\sigma_B\otimes\II_C +
\II_A\otimes\II_B\otimes\vec\sigma_C$ as the total spin operator,
a collective rotation operator of all three spins can be written
as $\exp\{-i\theta\vec n\cdot\vec J\}$. The map $\Phi$ is
therefore the statistical average of all such collective rotations
\begin{eqnarray}
\Phi[\rho] &=& \frac{1}{4\pi}\int_\S \exp\{-i\theta\vec n\cdot\vec J\} \rho \exp\{i\theta\vec n\cdot\vec J\} d\vec n \nonumber\\
&=& \frac{1}{3} \left[ E_x \rho E_x^\dagger +  E_y \rho
E_y^\dagger + E_z \rho E_z^\dagger \right]
\end{eqnarray}
where $E_k = \exp\{-i\theta J_k\}$, $k=x,y,z$. The second line
follows by the symmetry of the integration region.

Hence, the collective rotation channel is characterized by the
three angular momentum operators $J_x$, $J_y$, and $J_z$. The
noiseless subsystems for $\Phi$ are encoded in its ``noise
commutant''. This is the operator algebra $\A^\prime =
\{E_x,E_y,E_z\}^\prime = \{J_x,J_y,J_z\}^\prime$. Operators in the
noise commutant are fixed points for $\Phi$, and hence are immune
to the noise of $\Phi$. This algebra is unitarily equivalent to
the algebra $\A^\prime \cong {\mathbb C}\one_4 \oplus
(\one_2\otimes\M_2)$. Thus, the states encoded inside the
subalgebra isomorphic to $\one_2\otimes\M_2$ remain error-free
under $\Phi$, making use of symmetries in the noise. It is
important to note that the tensor structure determined by this
subalgebra is different than the initial system tensor
presentation determined by the three particles $A$, $B$ and $C$.
Let us be more specific.

The operators $J_x$, $J_y$, and $J_z$ form a representation of
$su(2)$, whose irreducible sectors are given by the eigenspaces of
the total angular momentum operator $J^2$. An orthonormal basis
for $\H_\S$ is $\ket{j,m,\mu}$, where $j = \frac 12,\ \frac 32$ is
the total spin number, where $J^2\ket{j,m,\mu} =
j(j+1)\ket{j,m,\mu}$, $m = -j,-j+1,\ldots j,$ is the projection of
the spin along the $z$ axis $J_z \ket{j,m,\mu} = m\ket{j,m,\mu}$,
and where $\mu$ labels the multiplicity of the representation.

From elementary composition of angular momentum, it can be found
that there is a single copy of the spin-$\frac 32$ representation
while the spin-$\frac 12$ representation appears in two copies.
Hence, in the subspace associated to the eigenvalue $j=\frac 12$,
$\H_{\frac 12}$, the states can be represented with two quantum
numbers $\ket{m,\mu}$, $m = \pm\frac 12$ labeling the eigenvalue
of $J_z$ and $\mu = 1,2$ labeling the two copies of the
irreducible representation. One can think of these two quantum
numbers as resulting from the tensor product of the Hilbert space
of two subsystems  $\H_{\frac 12} = \H_m\otimes\H_\mu$. The system
$\H_m$ gets completely mixed by the map $\Phi$ while the system
$\H_\mu$ is completely immune to noise. Thus, as discussed above,
any state of the form $\II_m \otimes \rho_\mu$ is a fixed point of
$\Phi$. More generally, Eq.~(\ref{ns}) is satisfied here since
$\Phi(\rho_m\otimes\rho_\mu) = \II_m \otimes \rho_\mu$.

Let us emphasize again that in this example, the division of the
Hilbert space $\H_\S = \H_{\frac 32} \oplus (\H_m\otimes\H_\mu)$
has no relation with its natural division $\H_\S =
(\mathbb{C}^2)^{\otimes 3}$ into three subsystems $A$, $B$, and
$C$. The noiseless subsystem $\H_\mu$ is an abstract construction
which involves all three spin-$\frac 12$ particles in a
non-trivial way. Furthermore, the subsystem $\H_m$ is virtually
absent from the dynamics as its state gets randomized by $\Phi$.
It is irrelevant to the physics of the system as it does not and
cannot convey information.



\end{document}